\documentclass[runningheads,a4paper]{llncs}

\makeatletter\let\proof\@undefined\let\endproof\@undefined\makeatother

\usepackage{bmpsize}
\usepackage{verbatim}
\usepackage{times}
\usepackage{fullpage}
\usepackage{xspace}
\usepackage{amsmath}
\usepackage{amsthm}
\usepackage{amssymb}
\usepackage{mathpartir}

\usepackage{semantic}
\usepackage{stmaryrd}
\usepackage{multirow}
\usepackage{graphicx}
\usepackage{listings}
\usepackage{color,xcolor,colortbl}
\usepackage{pgfplotstable}
\usepackage{outlines}
\usepackage{enumitem}
\usepackage{url}
\setlist{nosep}
\usepackage{url}
\usepackage{comment}
\usepackage{listings}
\usepackage{subfigure}
\usepackage{caption} 
\usepackage{upgreek}
\usepackage{pgfplots}
\usepackage{cleveref}
\usepackage{graphicx}
\usepackage{wrapfig}
\usetikzlibrary{patterns}
\usepackage{algorithm2e}
\LinesNumbered

\newif\iftr
\trfalse

\usepackage[space]{grffile}

\usepackage{tikz-qtree}
\usepackage{tikz}
\usetikzlibrary{arrows,calc,intersections,fit}
\usepackage{tikz-cd}

\tikzcdset{diagrams={tips=false,column sep=-.5em,row sep=1em,math mode=false}}

\newcommand{\typtodo}[1]{}
\newcommand{\dawn}[1]{{\color{blue}~{\textbf{[Dawn: #1]}}}}
\newcommand{\joe}[1]{{\color{brown}~{\textbf{[Joe: #1]}}}}
\newcommand{\ray}[1]{{\color{orange}~{\textbf{[Ray: #1]}}}}

\newcommand{\om}[1]{{\color{blue}~~~\emph{\textbf{#1}}}}
\newcommand{\lun}[1]{{\color{green}~{\textbf{[Lun: #1]}}}}
\newcommand{\muzhang}[1]{{\color{purple}~{\textbf{[Mu: #1]}}}}
\newcommand{\dan}[1]{{\color{purple}~{\textbf{[Dan: #1]}}}}
\newcommand{\ruoxi}[1]{{\color{blue}~{\textbf{[Ruoxi: #1]}}}}
\newcommand{\pgao}[1]{{\color{purple}~{\textbf{[Peng: #1]}}}}

\newif\ifcomments
\commentstrue

\ifcomments
\else
\renewcommand{\dawn}[1]{}
\renewcommand{\joe}[1]{}
\renewcommand{\ray}[1]{}
\renewcommand{\om}[1]{}
\renewcommand{\lun}[1]{}
\renewcommand{\muzhang}[1]{}
\renewcommand{\dan}[1]{}
\renewcommand{\ruoxi}[1]{}
\renewcommand{\pgao}[1]{}
\fi

\newcommand{\constraint}[1]{{\color{brown}{\noindent{[Constraint: #1]}}}}
\renewcommand{\constraint}[1]{}

\renewcommand{\paragraph}[1]{\vspace*{1mm}\noindent\textbf{#1}\phantom{b}}

\newif\ifsgx
\sgxfalse

\newif\ifextended
\extendedfalse

\ifsgx
\newcommand{\sgxcomment}[1]{#1}
\else
\newcommand{\sgxcomment}[1]{}
\fi

\ifsgx

\else

\fi

\lstdefinestyle{myScalastyle}{
  frame=tb,
  language=scala,
  aboveskip=2.5pt plus .5pt minus .5pt,
  belowskip=2.5pt plus .5pt minus .5pt,
  xleftmargin=4mm,
  xrightmargin=2mm,
  showstringspaces=false,
  columns=flexible,
  basicstyle={\small\ttfamily},
  numbers=none,
  numberstyle=\tiny\color{gray},
  keywordstyle=\color{blue},
  commentstyle=\color{gray},
  frame=none,
  breaklines=true,
  breakatwhitespace=true,
  tabsize=3,
}
\expandafter\def\expandafter\normalsize\expandafter{%
  \normalsize\setlength\abovedisplayskip{3pt}}
\expandafter\def\expandafter\normalsize\expandafter{%
  \normalsize\setlength\belowdisplayskip{4pt}}

\theoremstyle{remark}

\definecolor{Gray}{gray}{0.85}

\newcommand{\system}{\textsc{PrivGuard}\xspace}
\newcommand{\lang}{\textsc{PrivPolicy}\xspace}

\definecolor{light-gray}{gray}{0.95}

\tikzstyle{every picture}+=[remember picture]

\newcommand{\psep}{:}
\newcommand{\patt}[1]{\textsf{#1}}
\newcommand{\dc}{\mathbb{D}}
\newcommand{\legalease}{{\sc Legalease}\xspace}

\tikzset{every tree node/.style={minimum width=2em,draw,circle},
         blank/.style={draw=none},
         edge from parent/.style=
         {draw, edge from parent path={(\tikzparentnode) -- (\tikzchildnode)}},
         level distance=1.5cm}

\renewcommand{\paragraph}[1]{\vspace{1mm}\noindent\textbf{#1} }

\lstdefinelanguage{scala}{
	morekeywords={true,false,abstract,case,catch,class,def,%
		do,else,extends,false,final,finally,%
		for,if,implicit,import,match,mixin,%
		new,null,object,override,package,%
		private,protected,requires,return,sealed,%
		super,this,throw,trait,true,try,%
		type,val,var,while,with,yield,%
		Policy, deny, except, allow, Role},
	literate={epsilon}{$\epsilon$}{1} {datatype}{\textit{datatype }}{1} {column}{\textit{column }}{1} {runOn}{\textit{runOn }}{1}
	{accessBy}{\textit{accessBy }}{1} {where}{\textit{where }}{1} {using}{\textit{using }}{1},
	otherkeywords={=>,<-,<\%,<:,>:,\#,@},
	sensitive=true,
	morecomment=[l]{//},
	morecomment=[n]{/*}{*/},
	morestring=[b]",
	morestring=[b]',
	morestring=[b]""",
        breaklines=true,
	columns=flexible
}
\lstdefinelanguage{legalease}{
	keywords={deny,allow,except,Deny,Allow,Except,DENY,ALLOW,EXCEPT,AND,OR},
	keywordstyle=\bfseries,
	ndkeywords={Role,Policy,using,accessBy,datatype,column,where,runOn,
          Cond, Declass, Deployment, Column, Data, SuppressCol, Trans, schema, declass,
          ROLE, SCHEMA, DECLASS, FILTER, REDACT, PURPOSE, CONSENT_REQUIRED, NOTIFICATION_REQUIRED},
	ndkeywordstyle=\textit,
	literate={epsilon}{$\epsilon$}{1} {delta}{$\delta$}{1}
                 {section}{\S}{1}
                  {=>}{$\Rightarrow$}{1},
	sensitive=true,
	columns=flexible,
	numbers=left,
	stepnumber=1,
	firstnumber=1,
	numberfirstline=true,
	basicstyle=\small\sffamily,
}

\lstdefinelanguage{outline}{
        columns=flexible,
        basicstyle=\small\sffamily,
        breaklines=true
}

\lstnewenvironment{out-line}
    {\lstset{language=outline
    }}
    {}

\begin{document}

\mainmatter  

\title{Data Capsule: A New Paradigm for Automatic Compliance with Data Privacy Regulations}

\titlerunning{Lecture Notes in Computer Science: Authors' Instructions}

%
%
\author{Lun Wang$^\ddagger$ \and Joseph P. Near$^\dagger$ \and Neel Somani$^\ddagger$ \and
Peng Gao$^\ddagger$ \and Andrew Low$^\ddagger$ \and David Dao$^{\diamond}$ \and Dawn Song$^{\ddagger}$}


\institute{$\dagger$: University of Vermont
$\ddagger$: University of California, Berkeley
$\diamond$: ETH Zurich
}

%
%

\maketitle

\begin{abstract}
The increasing pace of data collection has led to increasing awareness of privacy risks, resulting in new data privacy regulations like General data Protection Regulation (GDPR). 
Such regulations are an important step, but automatic compliance checking is challenging. 
In this work, we present a new paradigm, \emph{Data Capsule}, for automatic compliance
checking of data privacy regulations in heterogeneous data processing infrastructures.
Our key insight is to pair up a data subject's data with a policy governing how the data is processed.
Specified in our formal policy language: \lang, the policy is created and provided by the data subject alongside the data, and is associated with the data throughout the life-cycle of data processing (e.g., data transformation by data processing systems, data aggregation of multiple data subjects' data).
We introduce a solution for static enforcement of privacy policies based on the concept of residual policies, and present a novel algorithm based on abstract interpretation for deriving residual policies in \lang.
Our solution ensures compliance automatically, and is designed for deployment alongside existing infrastructure. We also design and develop \system, a reference data capsule manager that implements all the functionalities of \emph{Data Capsule} paradigm.

\end{abstract}

\section{Introduction}

The big data revolution has triggered an explosion in the collection and processing of our personal data, leading to numerous societal benefits and sparking brand-new fields of research. At the same time, this trend of ever-increasing data collection raises new concerns about data privacy. The prevalence of data breaches~\cite{databreach}\cite{solove2017risk}, insider attacks~\cite{insiderattack}, and organizational abuses of collected data~\cite{murdock1979use} indicates that these concerns are well-founded. Data privacy has thus become one of the foundational challenges of today's technology landscape.




To address this growing challenge, governments have begun crafting data privacy regulations to protect us (the \emph{data subjects}) from those who collect and process our personal data. Recent examples include the European Union's \emph{General Data Protection Regulation} (GDPR)~\cite{GDPR}, the \emph{California Consumer Privacy Act} (CCPA)~\cite{CCPA}, the \emph{Family Educational Rights \& Privacy Act}~\cite{FERPA}, and the \emph{Health Insurance Portability and Accountability Act}~\cite{HIPAA}. 

Unfortunately, compliance with data privacy regulations is extremely challenging with current data processing systems. The regulations are written in natural language, and thus are difficult to formalize for automatic enforcement.
In addition, some of the systems currently used for data processing were designed and deployed before the existence of these privacy regulations, and their designs make the compliance even more difficult.
For example, many existing data processing systems do not provide an option to delete data, since it was assumed that organizations would want to keep data forever~\cite{forever}---but GDPR requires that a subject's data be deleted on request. Even if the deletion is possible, its enforcement can be challenging: organizations often make multiple copies of data, without no systematic record of the copies, because each data processing platform requires its own data format; as a result, an organization may not even be able to \emph{locate} all of the copies of a data subject's data for deletion \cite{maniatis2011you}.

Compliance with data privacy regulations is therefore costly or impossible for many organizations. These challenges reduce the rate of compliance, resulting in harm to data subjects via privacy violations. Moreover, the cost of implementing compliance acts as a barrier to entry for small organizations, and serves to protect large organizations from new competition. Paradoxically, new data privacy regulations may actually \emph{help} the large corporations whose abuses of data originally motivated those regulations.

This paper presents a new paradigm for \emph{automatic compliance} with data privacy regulations in heterogeneous data processing infrastructures. Our approach is based on a new concept called the \emph{data capsule}, which pairs up a data subject's data with a \emph{policy} governing how the data may be processed. The policy follows the data subject's data \emph{forever}, even when it is copied from one data processing system to another or mixed with data from other subjects. Our solution is designed for deployment \emph{alongside} existing infrastructure, and requires only minimal changes to existing data processing systems. The approach is \emph{automatic}, enabling compliance with minimal additional cost to organizations.

\paragraph{The Data Capsule Paradigm.}
We propose a new paradigm for collecting, managing, and processing sensitive personal data, called the Data Capsule Paradigm, which \emph{automates} compliance with data privacy regulations. Our paradigm consists of three major components:

\begin{enumerate}[topsep=1mm,leftmargin=4mm]
\itemsep0.5mm
\item \textbf{Data capsule}, which contains sensitive personal data, a \emph{policy} restricting how the data may be processed, and \emph{metadata} relevant for data privacy concerns.
\item \textbf{Data capsule graph}, which tracks all data capsules, including data collected from data subjects and data derived (via processing) from the collected data. 
\item \textbf{Data capsule manager}, which maintains the data capsule graph, registers new data capsules, enforces each capsule's policy, and propagates metadata through the graph.
\end{enumerate}

\paragraph{Principles of Data Privacy.}
To reach the design requirements for our solution, we examine four existing data privacy regulations (GDPR, CCPA, HIPAA, and FERPA). We propose five \emph{principles of data privacy} which accurately represent common trends across these regulations: \emph{transparency \& auditing}, \emph{consent}, \emph{processing control}, \emph{data portability}, and \emph{guarantee against re-identification}. 
Our principles are designed to be flexible. A solution targeting these principles can be made compliant with current data privacy regulations, and is also capable of being extended to new regulations which may be proposed in the future.

\paragraph{\lang: a Formal Privacy Policy Language.}
To enforce the five principles of data privacy outlined above, we introduce \lang: a novel \emph{formal policy language} designed around these principles, and capable of encoding the formalizable subset of recent data privacy regulations. By formalizable subset, we filter out requirements like ``legitimate business purpose'' in GDPR, which is almost impossible to formalize and have to rely on auditing to enforce requirements like this. We demonstrate the flexibility of \lang by encoding GDPR, CCPA, HIPAA, and FERPA.

\lang has a formal semantics, enabling a sound analysis to check whether a data processing program complies with the policy.
To enforce these policies, we present a novel static analysis based on abstract interpretation. The data capsule graph enables \emph{pipelines} of analysis programs which together satisfy a given policy. To enforce policies on these pipelines in a compositional way, we propose an approach which statically infers a \emph{residual policy} based on an analysis program and an input policy; the residual policy encodes the policy requirements which remain to be satisfied by later programs in the pipeline, and is attached to the output data capsule of the program.

Our approach for policy enforcement is entirely static. It scales to datasets of arbitrary size, and is performed as a pre-processing step (independent of the \emph{execution} of analysis programs). Our approach is therefore well-suited to the heterogeneous data processing infrastructures used in practice.

\paragraph{\system: a Data Capsule Manager.}
We design and implement \system, a reference data capsule manager. \system consists of components that manage the data capsule graph and perform static analysis of \emph{analysis programs} which process data capsules. \system is designed to work with real data processing systems and introduces negligible performance overhead. Importantly, \system makes no changes to the format in which data is stored or the systems used to process it. Its static analysis occurs as a separate step from the processing itself, and can be performed in parallel. In a case study involving medical data, we demonstrate the use of \system to enforce HIPAA in the context of analysis programs for a research study.

\paragraph{Contributions.} In summary, we make the following contributions.
\begin{itemize}[topsep=1mm,leftmargin=4mm]
\itemsep0.5mm
\item We propose five \emph{principles of data privacy} which encompass the requirements of major data privacy regulations.
\item We introduce \lang: a new and expressive formal language for privacy policies, which is capable of encoding policies for compliance with the formalizable subset of data privacy regulations.
\item We propose the data capsule paradigm, an approach for ensuring compliance with privacy regulations encoded using \lang, and formalize the major components of the approach.
\item We present the encoding of GDPR in \lang. 
\item We introduce a solution for static enforcement of privacy policies based on the concept of \emph{residual policies}, and present a novel algorithm based on abstract interpretation for deriving residual policies in \lang.
\item We design and develop \system, a reference data capsule manager that implements all the functionalities of data capsule paradigm.
\end{itemize}
\section{Requirements of Data Privacy Regulations}
\label{sec:data_privacy}

Recent years have seen new efforts towards regulating data privacy, resulting in regulations like the European Union's \emph{General Data Protection Regulation} (GDPR). It joins more traditional regulations like the \emph{Health Insurance Portability and Accountability Act} (HIPAA) and the \emph{Family Educational Rights and Privacy Act} (FERPA).




\subsection{Principles of Data Privacy}

Historically, organizations have collected as much personal data as possible, and have not generally considered data privacy to be a high priority. The recent adoption of GDPR has forced a much wider set of organizations to consider solutions for ensuring compliance with data privacy regulations. Complying with regulations like GDPR is extremely difficult using existing systems, which generally are designed for \emph{easy access to data} instead of strong data privacy protections. These regulations are even more difficult to satisfy when data is shared between organizational units or with third parties---yet the regulatory requirements apply even in these cases.

To address this challenge, we considered the \emph{commonalities} between the three major data privacy regulations mentioned above to develop five \emph{principles of data privacy}. These principles expose and generalize the fundamental ideas shared between regulations, and therefore are likely to also apply to future regulations. As described in Section~\ref{sec:data-caps-parad}, these five principles form the design criteria for our proposed data capsule paradigm.

In describing the five principles of data privacy, we use terminology from the GDPR.
The term \emph{data subject} refers to individuals whose data is being collected and processed, and the term \emph{data controller} refers to organizations which collect and process data from data subjects. Briefly summarized, the five principles of data privacy are:

\begin{enumerate}[topsep=1mm,leftmargin=4mm]
\itemsep1.0mm
\item \textbf{Transparency \& Auditing:} The data subject should be made aware of who has their data and how it is being processed.
\item \textbf{Consent:} The data subject should give explicit consent for the collection and processing of their data.
\item \textbf{Processing Control:} The data subject should have control over what types of processing are applied to their data.
\item \textbf{Data Portability:} The data subject should be able to obtain a copy of any data related to them.
\item \textbf{Guarantee Against Re-identification:} When possible, the results of processing should not permit the re-identification of any individual data subject.
\end{enumerate}

\subsection{Applying the Principles}

The five principles described above represent the design criteria for our data capsule paradigm. They are specified specifically to be \emph{at least as strong as} the common requirements of existing privacy regulations, to ensure that our approach is capable of expressing a large enough subset of existing and future regulations to ensure compliance. This section demonstrates how our five principles describe and subsume the requirements of the four major privacy regulations.

\paragraph{GDPR.}
The major pillars of GDPR fall squarely into the five requirement categories described by our principles. Articles 13 and 14 describe \emph{transparency \& auditing} requirements: the data controller must inform the data subject about the data being collected and who it is shared with. Article 4, 7 and 29WP requires \emph{consent}: the controller must generally obtain consent from the data subject to collect or process their data. Note that there are also cases in GDPR when personal data can be used without consent, where some other ``lawful basis for processing'' applies, such as public interest, legal obligation, contract or the legitimate interest of the controller. However, these purposes are almost impossible to formalize so we have to rely on auditing to enforce them and omit them in the system. Articles 18 and 22 ensure \emph{processing control}: the data subject may allow or disallow certain types of processing. Articles 15, 16, 17, and 20 require \emph{data portability}: the data subject may obtain a copy of their data, fix errors in it, and request that it be deleted. Finally, Recital 26 and Article 29WP describes a \emph{guarantee against re-identification}: data controllers are required to take steps to prevent the re-identification of data subjects.

\paragraph{CCPA.}
CCPA is broadly similar to GDPR, with some differences in the specifics. Like GDPR, the requirements of CCPA align well with our five principles of data privacy. Unlike GDPR, CCPA's \emph{consent} requirements focus on the \emph{sale} of data, rather than its original collection. Its \emph{access \& portability} requirements focus on data deletion, and are more limited than those of GDPR. Like GDPR, CCPA ensures a \emph{guarantee against re-identification} by allowing data subjects to sue if they are re-identified.

\paragraph{HIPAA.}
The HIPAA regulation is older than GDPR, and reflects a contemporaneously limited understanding of data privacy risks. HIPAA requires the data subject to be notified when their data is collected (\emph{transparency \& auditing}), and requires \emph{consent} in some (but not all) cases. HIPAA requires organizations to store data in a way that prevents its unintentional release (partly ensuring a \emph{guarantee against re-identification}), and its ``safe harbor'' provision specifies a specific set of data attributes which must be redacted before data is shared with other organizations (an attempt to ensure a \emph{guarantee against re-identification}). HIPAA has only limited \emph{processing control} and \emph{data portability} requirements.

\paragraph{FERPA.}
The \emph{Family Educational Rights and Privacy Act of 1974} (FERPA) is a federal law in the United States that protects the privacy of student education records. FERPA requires \emph{consent} before a post-secondary institution shares information from a student's education record. It also requires \emph{access \& portability}: students may inspect and review their records, and request amendments. In other respects, FERPA has fewer requirements than the other regulations described above.
\vspace{-0.1in}
\section{The Data Capsule Paradigm}
\label{sec:data-caps-parad}

This section introduces the \emph{data capsule paradigm}, an approach for addressing the five principles of data privacy described earlier. The data capsule paradigm comprises four major concepts:

\begin{itemize}[topsep=1mm,leftmargin=4mm]
\itemsep0.5mm
\item \emph{Data capsules} combine sensitive data contributed by a data subject (or derived from such data) with a \emph{policy} governing its use and \emph{metadata} describing its properties.
\item \emph{Analysis programs} process the data stored inside data capsules; the input of an analysis program is a set of data capsules, and its output is a new data capsule.
\item The \emph{data capsule graph} tracks all data capsules and analysis programs, and contains edges between data capsules and the analysis programs which process them.
\item The \emph{data capsule manager} maintains the data capsule graph, propagates policies and metadata within the graph, and controls access to the data within each data capsule to ensure that capsule policies are never violated.
\end{itemize}

In Section~\ref{sec:satisfying}, we demonstrate how these concepts are used to satisfy our five principles of data privacy. Section~\ref{sec:system:-data-capsule} describes \system, our proof-of-concept data capsule manager which implements the paradigm.

\vspace{-0.1in}
\subsection{Life Cycle of a Data Capsule}
\label{sec:lifecycle}

\begin{wrapfigure}{r}{0.4\textwidth}
\centering
\includegraphics[width=.3\textwidth]{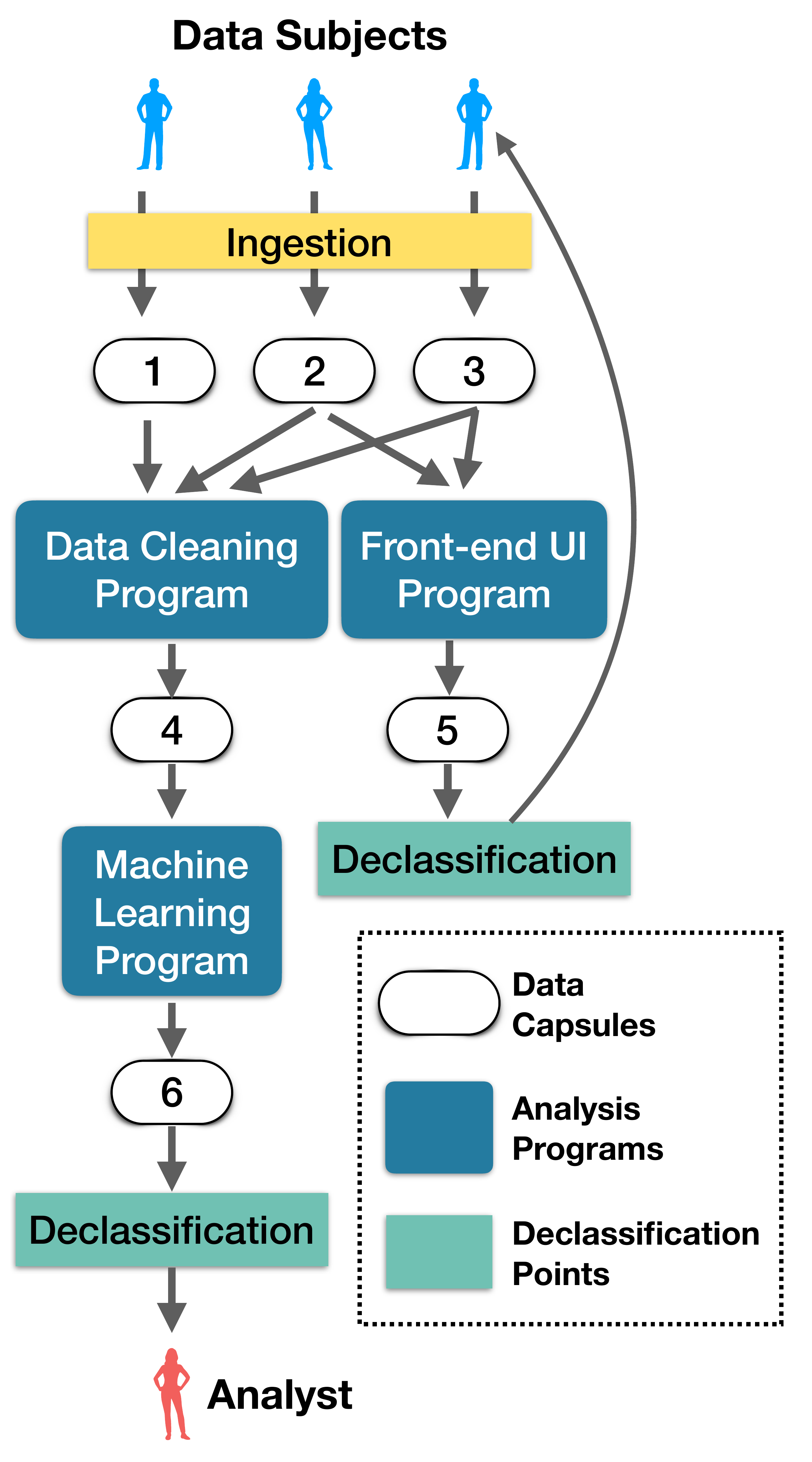}
\caption{Example Data Capsule Graph}
\label{fig:dcg_example}
 \vspace{-0.2in}
\end{wrapfigure}

The life cycle of a data capsule includes four phases:
\begin{enumerate}[topsep=1mm,leftmargin=4mm]
\itemsep0.5mm
\item \textbf{Data Ingestion.}
\emph{Data subjects} construct data capsules from their sensitive data via the \emph{ingestion} process, which pairs the data with the policy which will govern its use. In our setting, the data subject is the original data owner, whose privacy we would like to protect.

\item \textbf{Analysis Program Submission.}
\emph{Analysts} who would like to process the data contained in data capsules may submit \emph{analysis programs}, which are standard data analytics programs augmented with API calls to the data capsule manager to obtain raw data for processing.

\item \textbf{Data Processing.}
  Periodically, or at a time decided by the analyst, the data capsule manager may run an analysis program. At this time, the data capsule manager statically determines the set of input data capsules to the program, and performs static analysis to verify that the program would not violate the policies of any of its inputs. As part of this process, the data capsule manager computes a \emph{residual policy}, which is the new policy to be attached to the program's output. The data capsule manager then runs the program, and constructs a new data capsule by pairing up the program's output with the residual policy computed earlier.

\item \textbf{Declassification.}
A data capsule whose policy has been satisfied completely may be viewed by the analyst in a process called \emph{declassification}. When an analyst requests that the data capsule manager declassify a particular data capsule, the manager verifies that its policy has been satisfied, and that the analyst has the appropriate role, then sends the raw data to the analyst. Declassification is the only process by which data stored in a data capsule can be divorced from its policy.
\end{enumerate}

\vspace{-0.1in}
\subsection{The Data Capsule Manager}

The data capsule life cycle is supported by a system implementing the functionality of the data capsule manager. The primary responsibility of the data capsule manager is to maintain the data capsule graph and maintain its invariants---namely, that no data capsule's policy is violated, that new data capsules resulting from analysis programs have the right policies, and that metadata is propagated correctly. We describe our reference implementation, \system, in Section~\ref{sec:system:-data-capsule}.

Figure~\ref{fig:dcg_example} contains a global view of an example data capsule graph. This graph contains two different organizations representing data controllers, and data subjects associated with each one. Both organizations use analysis programs which combine and clean data from multiple data subjects into a single data capsule; a third analyst uses data capsules from \emph{both} organizations to perform marketing research. Such a situation is allowed under privacy regulations like GDPR, as long as the policies specified by the data subjects allow it. This example therefore demonstrates the ability of the data capsule paradigm to break down data silos while at the same time maintaining privacy for data subjects---a key benefit of the paradigm.

In this example, the policy attached to each data subject's capsule is likely to be a formal representation of GDPR. The data capsule paradigm requires a formal encoding of policies with the ability to efficiently compute residual policies; we describe our solution to this challenge in Section~\ref{sec:policies}.

Note that the data capsules containing the data subjects' combined data (capsules 1, 2, 3, and 4) cannot be viewed by \emph{anyone}, since their policies have not been satisfied. This is a common situation in the data capsule paradigm, and it allows implementing useful patterns such as extract-transform-load (ETL) style pipelines~\cite{ETL}. In such cases, analysts may submit analysis programs whose primary purpose is to prepare data for \emph{other} analysis programs; after being processed by some (potentially long) pipeline of analysis programs, the final output has satisfied all of the input policies and may be declassified. The intermediate results of such pipelines can never be viewed by the analyst.

\vspace{-0.1in}
\subsection{Satisfying the Principles of Data Privacy}
\label{sec:satisfying}

  
The data capsule paradigm is designed specifically to enable systems which satisfy the principles of data privacy laid out in Section~\ref{sec:data_privacy}.

\paragraph{Transparency \& Auditing.} The data capsule manager satisfies {transparency \& auditing} by consulting the data capsule graph. The global view of the graph (as seen in Figure~\ref{fig:dcg_example}) can be restricted to contain only the elements reachable from the ingested data capsules of a single data subject, and the resulting sub-graph represents {all of the data collected about or derived from the subject}, plus {all of the processing tasks performed} on that data.


\paragraph{Consent.} The data capsule manager tracks consent given by the data subject as metadata for each data capsule. Data subjects can be prompted to give consent when new analysis programs are submitted, or when they are executed.

\paragraph{Processing Control.} The formal policies attached to data capsules can restrict the processing of the data stored in those capsules. These policies typically encode the restrictions present in data privacy regulations, and the data capsule manager employs a static analysis to verify that submitted analysis programs do not violate the relevant policies. This process is described in Section~\ref{sec:policies}.

\paragraph{Data Portability.} To satisfy the data portability principle, the data capsule manager allows each data subject to download his or her data capsules. The data capsule manager can also provide data capsules \emph{derived} from the subject's capsules, since these are reachable capsules in the data capsule graph. However, the derived data returned to the data subject must \emph{not} include data derived from the capsules of \emph{other subjects}, so a one-to-one mapping must exist between rows in the input and output capsules for each analysis program involved. We formalize this process in Section~\ref{sec:policies}.

The same mechanism is used for data deletion. When a data subject wishes to delete a capsule, the set of capsules derived from that capsule is calculated, and these derived capsules are re-computed \emph{without} the deleted capsule included in the input.

\paragraph{Guarantee Against Re-identification.} To provide a robust formal guarantee against re-identification, the data capsule manager supports the use of various techniques for anonymization, including both informal techniques (e.g. removing ``personal health information'' to satisfy HIPAA) and formal techniques (e.g. $k$-anonymity, $\ell$-diversity, and differential privacy). A data capsule's policy may require that analysis programs apply one of these techniques to protect against re-identification attacks. 
\section{\system: a Data Capsule Manager}
\label{sec:system:-data-capsule}

We have designed and implemented a reference data capsule manager, called \system. The \system system manages the data capsule graph, propagates policies and metadata, and uses static analysis to calculate residual policies on analysis programs.

\begin{wrapfigure}{r}{0.5\textwidth}
  \centering
  \includegraphics[width=.45\textwidth]{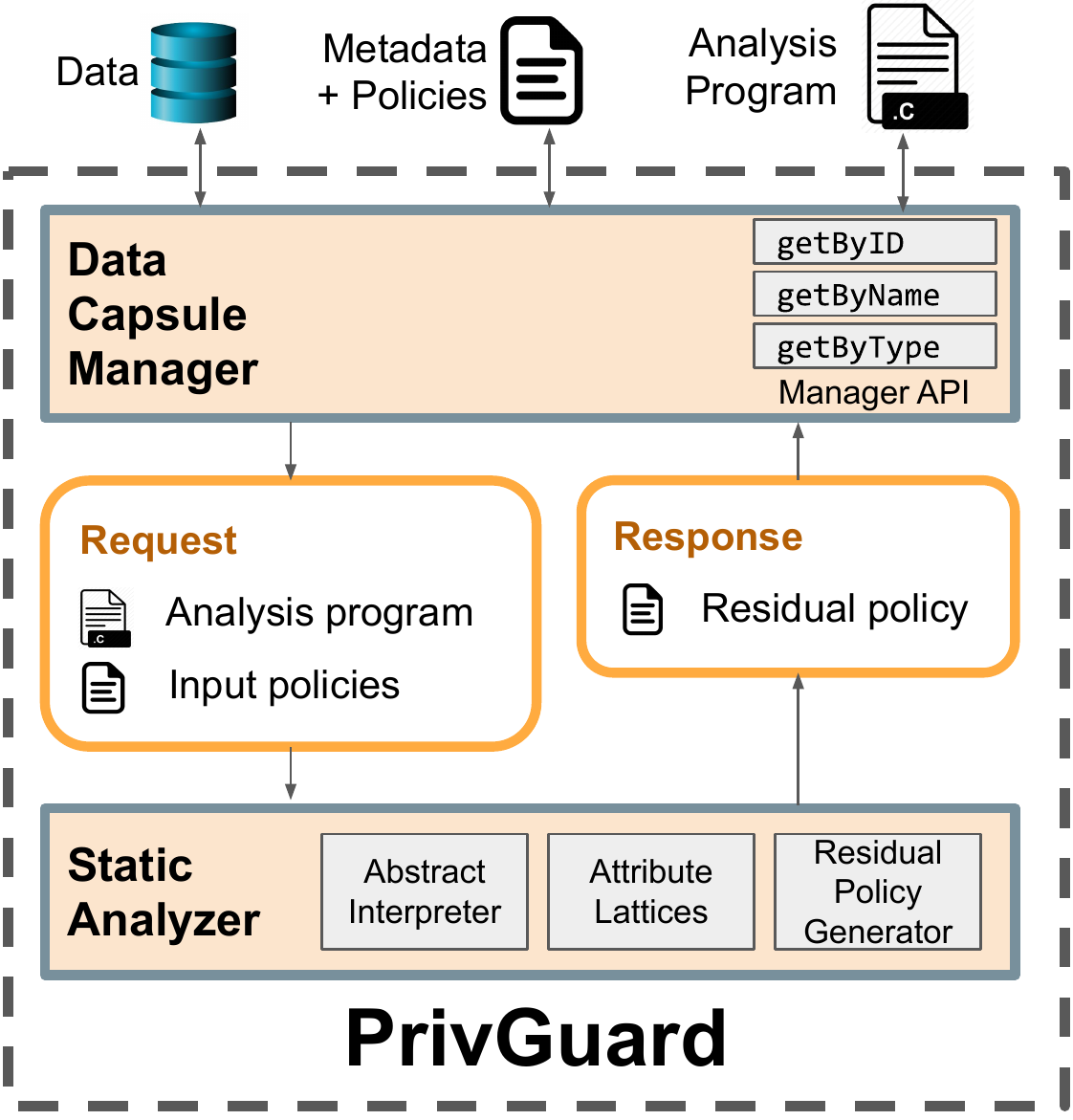}
  \caption{The Architecture of \system.}
  \label{fig:architecture}
  \vspace{-0.3in}
\end{wrapfigure}

Figure~\ref{fig:architecture} summarizes the architecture of \system. The two major components of the system are the data capsule manager itself, which maintains the data capsule graph, and the static analyzer, which analyzes policies and analysis programs to compute residual policies. We describe the data capsule manager here, and formalize the static analyzer in Section~\ref{sec:policies}.

Finally, \lstinline|outputCapsule| defines an output data capsule of the analysis program. The analyst specifies a dataframe containing the output data, and \system automatically attaches the correct residual policy. This process is formalized in Section~\ref{sec:policies}.

\paragraph{Deployment \& Integration.}
The data capsule paradigm is intended to be integrated with existing heterogeneous data processing infrastructures, like the ones already in place for data analysis at many organizations, and \system is designed to facilitate such deployments. These infrastructures leverage a variety of data stores, including SQL databases~\cite{sql}, key/value stores like MongoDB~\cite{mongodb}, distributed filesystems like HDFS~\cite{hdfs}, and short-term publish/subscribe systems like Cassandra~\cite{cassandra}. They employ many different techniques for processing the data, including SQL engines and distributed systems like MapReduce~\cite{mapreduce}, Hadoop~\cite{hadoop}, and Spark~\cite{spark}.

To work successfully in such a heterogeneous environment, \system is deployed \emph{alongside} the existing infrastructure. As shown in Figure~\ref{fig:architecture}, policies and metadata are stored separately from the data itself, and the data can remain in the most efficient format for processing (e.g. stored in CSV files, in HDFS, or in a SQL database).

Similarly, \system's static analyzer uses a common representation to encode the semantics of many different kinds of analysis programs, so it works for many programming languages and platforms. The only platform-specific code is the small \system API, which allows analysis programs to interact with the data capsule manager. Our static analysis is based on abstract interpretation, a concept which extends to all common programming paradigms. Section~\ref{sec:policies} formalizes the analysis for dataflow-based systems which are close to relational algebra (e.g. SQL, Pandas, Hadoop, Spark); extending it to functional programs or traditional imperative or object-oriented programs is straightforward.
\vspace{-0.15in}
\section{Policies \& Policy Enforcement}
\label{sec:policies}
\vspace{-0.1in}

This section describes our formal language for specifying policies on
data capsules, and our static approach for enforcing these policies
when an analytics program is registered with the system. We describe
each of the four major components of this approach:

\begin{itemize}
\item Our policy specification language: \lang (\S~\ref{sec:policy-spec-lang}).
\item A set of {attribute} definitions suitable for encoding
  policies like GDPR and HIPAA, which are more expressive than the
  corresponding attributes proposed in previous work
  (\S~\ref{sec:policy-attributes}).
\item A flexible approach for deriving the \emph{policy effects} of an
  analysis program via abstract interpretation
  (\S~\ref{sec:abstr-interpr-analys}).
\item A formal procedure for determining the \emph{residual policy} on
  the output of an analysis program (\S~\ref{sec:comp-resid-polic}).
\end{itemize}

\subsection{\lang: Policy Specification Language}
\label{sec:policy-spec-lang}

\begin{table}
\centering
\begin{tabular}{ccc}
\begin{tabular}{rclcl}
$A$ & $\in$ & $\text{attribute}$ & $::=$ & $\mathit{attrName}\; \mathit{attrValue}$\\
$C$ & $\in$ & $\text{policy~clause}$ & $::=$ & $A \mid A\; \mathsf{AND}\; C \mid A\; \mathsf{OR}\; C$ \\
$P$ & $\in$ & $\text{policy}$ & $::=$ & $(\mathsf{ALLOW}\; C)^{+}$ \\
\end{tabular}
& \quad\quad & 
\begin{tabular}{rcl}
$A$     & $::=$       & $\mathit{attrName}\; \mathit{attrValue}$\\
$C_{DNF}$ & $\subseteq$ & $\mathcal{P}(A)$ \\
$P_{DNF}$ & $\subseteq$ & $\mathcal{P}(C_{DNF})$ \\
\end{tabular}\\
\\
    (1) \lang surface syntax. & & (2) \lang disjunctive normal form.
\end{tabular}
\caption{Surface Syntax \& Normal Form.}
\label{fig:syntax}
\end{table}

Our policy specification language: \lang is inspired by the \legalease
language~\cite{sen2014bootstrapping}, with small changes to surface syntax to
account for our more expressive attribute lattices and ability to
compute residual policies. 
\begin{wrapfigure}{l}{0.5\textwidth}
\begin{lstlisting}[language=legalease]
ALLOW SCHEMA NotPII
  AND NOTIFICATION_REQUIRED
  AND (ROLE $user_id
     OR (CONSENT_REQUIRED
          AND DECLASS DP 1 0.000001))
\end{lstlisting}
\caption{A subset of GDPR using \lang.}
\label{fig:hipaa_example1}
\vspace{-0.3in}
\end{wrapfigure}
The grammar for the surface syntax of \lang is given in Figure~\ref{fig:syntax} (1). The
language allows specifying an arbitrary number of \emph{clauses}, each
of which encodes a formula containing conjunctions and disjunctions
over attribute values. Effectively, each clause of a policy in our
language encodes one way to satisfy the overall policy.

\paragraph{Example.}
Figure~\ref{fig:hipaa_example1} specifies a subset of GDPR using \lang.
Each \lstinline[language=legalease]{ALLOW} keyword denotes a clause of the policy, and
\lstinline[language=legalease]{SCHEMA},
\lstinline[language=legalease]{NOTIFICATION_REQUIRED},
\lstinline[language=legalease]{ROLE},
\lstinline[language=legalease]{CONSENT_REQUIRED}, and
\lstinline[language=legalease]{DECLASS}
are attributes. This subset includes only a single clause, which says that information which is not personally identifiable may be processed by the data controller, as long as the data subject is notified, and either the results are only viewed by the data subject, or the data subject gives consent and differential privacy is used to prevent re-identification based on the results.


\begin{wrapfigure}{r}{0.5\textwidth}
\vspace{-0.4in}
{\small
\begin{align*}
\{ \{ & \textit{SCHEMA}\; \patt{NotPII}, \textit{NOTIFICATION\_REQUIRED},\\
      & \textit{ROLE}\; \patt{\$user\_id}\}\\
   \{ & \textit{SCHEMA}\; \patt{NotPII}, \textit{NOTIFICATION\_REQUIRED},\\
      & \textit{CONSENT\_REQUIRED}, \textit{DECLASS}\; \patt{DP(1.0)} \} \}\\
\end{align*}
}
\vspace{-0.4in}
\caption{Disjunctive normal form of the example policy.}
\label{fig:hipaa_example2}
\end{wrapfigure}
\paragraph{Conversion to Disjunctive Normal Form.}
Our first step in policy enforcement is to convert the policy to
\emph{disjunctive normal form} (DNF), a common conversion in
constraint solving. Conversion to DNF requires removing $\mathsf{OR}$
expressions from each clause of the policy; we accomplish this by
distributing conjunction over disjunction and then splitting the
top-level disjuncts within each clause into separate clauses. After
converting to DNF, we can eliminate the explicit uses of \lstinline[language=legalease]{AND} and \lstinline[language=legalease]{OR}, and represent the policy as a
\emph{set of clauses}, each of which is a \emph{set of attributes} as shown in Figure~\ref{fig:syntax} (2). The disjunctive normal form of our running example policy is shown in Figure~\ref{fig:hipaa_example2}. Note that the disjunctive normal form of our example contains two clauses, due to the use of \lstinline[language=legalease]{OR} in the original policy.

\vspace{-0.15in}
\subsection{Policy Attributes}
\label{sec:policy-attributes}
\vspace{-0.1in}

\legalease~\cite{sen2014bootstrapping} organizes attribute values into \emph{concept
  lattices}~\cite{conceptlattice}, and these lattices give policies their semantics. Instead of concept lattices, \lang leverages 
\emph{abstract domains} inspired by work on abstract
interpretation of programs~\cite{nielson2015principles}. This novel approach enables more expressive attributes (for example, the \lstinline[language=legalease]{FILTER} attribute) and also formalizes the connection between the semantics of policies and the semantics of analysis programs.

We require each attribute domain to define the standard lattice
operations required of an abstract domain: a partial order ($\sqsubseteq$), join ($\sqcup$), and meet
($\sqcap$), as well as top and bottom elements $\top$ and $\bot$. Many of these can be defined in terms of the
corresponding operations of an existing abstract domain from the abstract interpretation literature.




\paragraph{Filter Attributes.}
One example of our expressive attribute domains is the one for the
$\textit{FILTER}$ attribute, which
filters data based on integer-valued fields. The attribute domain for
$\textit{FILTER}$ is defined in terms of an \emph{interval} abstract
domain~\cite{nielson2015principles}. 
We say $\patt{filter} \psep f \psep i$
when the value of column $f$ lies in the interval $i$. Then, we define the following operations on $\textit{FILTER}$
attributes, completing its attribute domain:

\begin{equation}
\begin{array}{rclcl}
  \patt{filter} \psep f \psep i_1 &\sqcup& \patt{filter} \psep f \psep i_2 &=&
  \patt{filter} \psep f \psep i_1 \sqcup i_2\\
  \patt{filter} \psep f \psep i_1 &\sqcap& \patt{filter} \psep f \psep i_2 &=&
  \patt{filter} \psep f \psep i_1 \sqcap i_2\\
  \patt{filter} \psep f \psep i_1 &\sqsubseteq& \patt{filter} \psep f \psep i_2 &=&
  \psep i_1 \sqsubseteq i_2\\
\end{array}
\end{equation}

\paragraph{Schema Attributes.}
The schema attribute leverages a \emph{set} abstract domain, in which
containment is defined in terms of an underlying (finite) 
lattice of datatypes:

\begin{equation}
\begin{array}{rclcl}
  \patt{schema} \psep S_1 &\sqcup& \patt{schema} \psep S_2 &=& \patt{schema} \psep \{s' \mid s_1 \in S_1 \wedge s_2 \in S_2 \wedge s' = s_1 \sqcup s_2 \}  \\
  \patt{schema} \psep S_1 &\sqcap& \patt{schema} \psep S_2 &=& \patt{schema} \psep \{s' \mid s_1 \in S_1 \wedge s_2 \in S_2 \wedge s' = s_1 \sqcap s_2 \}  \\
  \patt{schema} \psep S_1 &\sqsubseteq& \patt{schema} \psep S_2 &=&  \forall s_1 \in S_1 , s_2 \in S_2 \;.\; s_1 \sqsubseteq s_2   \\
\end{array}
\end{equation}




\paragraph{Other Attributes.}
In \lang, as in \legalease, the partial ordering for
analyst roles is typically finite. It encodes the important properties
of each analyst (e.g. for GDPR, the government typically has more authority
to analyze data than members of the public).
The $\patt{role}$, $\patt{declass}$, and $\patt{redact}$ attributes
are defined by finite lattices. We omit the details here.







\subsection{Abstract Interpretation of Analysis Programs}
\label{sec:abstr-interpr-analys}

\begin{figure*}[h]
$$
f \in \mathsf{field} \quad
m \in \mathsf{int} \quad
s \in \mathsf{schema} \quad
x \in \mathsf{data~capsules}
$$
$$
\begin{array}{rclcl}
\delta & \in & \mathsf{filter} & ::= & f < m \mid f > m  \\
e & \in & \mathsf{expression} & ::= & \mbox{getDC}(x) \mid \mbox{filter}(\varphi, e) \mid \mbox{project}(s, e) \mid \mbox{redact}(a, e) \\
 &  &  & \mid & \mbox{join}(e, e) \mid \mbox{union}(e,e) \mid \mbox{dpCount}(\epsilon, \delta, e) \\
\end{array}
$$
\caption{Program Surface Syntax}
\label{fig:simple_syntax}
\end{figure*}

We next describe the use of abstract interpretation to determine the
\emph{policy effect} of an analysis program. We introduce this concept
using a simple dataflow-oriented language, similar to relational
algebra, Pandas, or Spark, presented in
Figure~\ref{fig:simple_syntax}.
We write an abstract data capsule with schema $s$ and policy effect
$\psi$ as $\dc[s, \psi]$. A data capsule environment $\Delta$ maps
data capsule IDs to their schemas (i.e. $\Delta : id \rightarrow s$).

\begin{figure*}
\begin{mathpar}
\inferrule*[Right=GetDC]
{\Delta(id) = s}
{\Delta \vdash \mbox{getDC}(id) : \dc[s, \emptyset]}

\and

\inferrule*[Right=Filter]
{\Delta \vdash e : \dc[s, \psi]
\and \varphi \rightsquigarrow_s a:v}
{\Delta \vdash \mbox{filter}(\varphi, e) : \dc[s, \psi + \patt{filter} \psep a:v]}

\and

\inferrule*[Right=Project]
{\Delta \vdash e : \dc[s, \psi]
\and s' \subseteq s}
{\Delta \vdash \mbox{project}(s', e) : \dc[s', \psi + \patt{schema} \psep s']}

\and

\inferrule*[Right=Redact]
{\Delta \vdash e : \dc[s, \psi]
\and a \in s
\and e_r \rightsquigarrow_s v}
{\Delta \vdash \mbox{redact}(a, e_r, e) : \dc[s, \psi + \patt{redact} \psep a:v]}

\and

\inferrule*[Right=Join]
{\Delta \vdash e_1 : \dc[s_1, \psi_1]
\and \Delta \vdash e_2 : \dc[s_2, \psi_2]}
{\Delta \vdash \mbox{join}(e_1, e_2) : \dc[s_1 \cup s_2, \psi_1 \cup \psi_2]}

\and

\inferrule*[Right=Union]
{\Delta \vdash e_1 : \dc[s, \psi_1]
\and \Delta \vdash e_2 : \dc[s, \psi_2]}
{\Delta \vdash \mbox{union}(e_1, e_2) : \dc[s, \emptyset]}

\and

\inferrule*[Right=DPCount]
{\Delta \vdash e : \dc[s, \psi]}
{\Delta \vdash \mbox{dpCount}(\epsilon, \delta, e) : \dc[s, \psi + \patt{declass} \psep \patt{DP}(\epsilon, \delta)]}

\end{mathpar}

\caption{Sample rules implementing an abstract interpreter for the data capsule expressions in the language presented in Figure~\ref{fig:simple_syntax}.}
\label{fig:collecting_semantics}
\end{figure*}

We present the \emph{abstract interpreter}~\cite{nielson2015principles} for \lang in Figure~\ref{fig:collecting_semantics}. If we can use the
semantics to build a derivation tree of the form $\Delta \vdash e :
\dc[s, \psi]$, then we know that the program is guaranteed to satisfy
the policy clause $\psi$ (or any clause which is less restrictive than
$\psi$).

\subsection{Computing Residual Policies}
\label{sec:comp-resid-polic}

Let $\Upsilon(id)$ be the policy of the data capsule with ID $id$. The
free variables of a program $e$, written $\textit{fv}(e)$, are the
data capsule IDs it uses.

We define the \emph{input policy} of a program $e$ to be the least
upper bound of the policies of its free variables:

\[ \Upsilon_{\patt{in}(e)} = \bigsqcup_{id \in \textit{fv}(e)} \{\Upsilon(id)\} \]

This semantics means that the input policy will be \emph{at least as
  restrictive} as the \emph{most restrictive} policy on an input data
capsule. It is computable as follows, because the disjunctive normal
form of a policy is a set of sets:

\[ p_1 \sqcup p_2 = \{ c_1 \cup c_2 \mid c_1 \in p_1 \wedge c_2 \in p_2 \} \]

The \emph{residual policy} applied to the output data capsule is
computed by considering each clause in the input policy, and computing
its residual based on the policy effect of the program. The residual
policy is computed using the following rule:

\begin{mathpar}
\inferrule*[Right=RP]
{\vdash e : \dc[s, \psi] }
{\Upsilon_{\patt{out}(e)} = \{ c' \mid c \in \Upsilon_{\patt{in}(e)} \wedge \mathit{residual}(c, \psi) = c' \} }
\end{mathpar}

where

\begin{align*}
\mathit{residual}(c, \psi) &= c - \{ k \psep p \mid k \psep p \in c \wedge \mbox{satisfies}(k \psep p, \psi) \}\\
\mathit{satisfies}(k \psep p, \psi) &= \exists k \psep p' \in \psi . p \sqsubseteq p'\\
\end{align*}

Here, the $\mathit{satisfies}$ relation holds for an attribute $k
\psep p$ in the policy when there exists an attribute $k \psep p'$ in
the policy effect of the program, such that $p$ (from the policy) is
less restrictive than $p'$ (the guarantee made by the
program). Essentially, we compute the residual policy from the input
policy by \emph{removing} all attributes for which
$\mathit{satisfies}$ holds.

\section{Formally Encoding GDPR}


\begin{wrapfigure}{r}{0.6\textwidth}
\vspace{-0.3in}
\begin{lstlisting}[language=legalease]
ALLOW SCHEMA NotPII
  AND NOTIFICATION_REQUIRED
  AND (ROLE $user_id
     OR (CONSENT_REQUIRED
         AND DECLASS DifferentialPrivacy 1 0.000001))

# Definitions mainly given in Article 6. Also Article 4, 25, 32
ALLOW SCHEMA PersonalInformation (Article 9)
  AND CONSENT_REQUIRED (Article 4, 6)

ALLOW SCHEMA PersonalInformation (Article 9)
  AND ROLE UserAffiliatedOrganizations($user_id)
  AND SCHEMA HasAppropriateSafeguards (article 25, 32, 46)

ALLOW SCHEMA PersonalInformation
  AND ROLE SupervisoryAuthority OR ROLE HealthcareOrganization
  AND PURPOSE PublicInterest LegalObligation PublicHealth 

ALLOW SCHEMA PersonalInformation
  AND ROLE LegalAuthority
  AND PURPOSE PublicInterest ForJudicialPurposes
\end{lstlisting}
\caption{Formal encoding of GDPR.}
\label{fig:GDPR}
\vspace{-0.3in}
\end{wrapfigure}

This section describes our formal encoding of a subset of GDPR, which is intended to ensure automated compliance with the regulation. We are in the process of developing similar encodings for other regulations, including HIPAA, FERPA.


Figure~\ref{fig:GDPR} contains our formal encoding. The first clause (lines 1-5) allows the use of data for any purpose, as long as it is protected against re-identification and subject to consent by the data subject. The third clause (lines 11-13) allows the use of personal information by organizations affiliated with the data subject---a relationship which we encode as a metafunction. The final two clauses specify specific public interest exceptions, for public health (lines 15-17) and for judicial purposes (lines 19-21).

GDPR is designed specifically to be simple and easy for users to understand, and its requirements are well-aligned with our five principles of data privacy. Our formal encoding is therefore correspondingly simple. We expect that most uses of data will fall under the third clause (for ``business uses'' of data, e.g. displaying Tweets to a Twitter user) or the first clause (for other purposes, e.g. marketing).

Data subjects who wish to modify this policy will generally specify more rigorous settings for the technologies used to prevent their re-identification. For example, a privacy-conscious data subject may require differential privacy in the first clause, instead of allowing any available de-identification approach.

\section{Performance Evaluation}
\label{sec:eval}

Recall that one of the goals of \system is to easily work with existing heterogeneous data processing systems and incur smallest additional overhead to the analysis itself. In order to achieve this goal, \system must have good scalability when the number of users are large. In this section, we first conduct an end-to-end evaluation to figure the bottleneck of the system scalability. Then we conduct several experimental evaluation to test the scalability of the bottleneck component. Specifically, we want to answer the following question: how is the scalability of \system and how much overhead \system will incur into the original data processing system.

\subsection{Experimental Design \& Setup} 

We first conduct an end-to-end evaluation to determine the sources of overhead when \system is used in a complete analysis. We then focus on the performance of policy ingestion as described in Section~\ref{sec:lifecycle}, which turns out to be the largest source of overhead in \system.  We vary the number of data capsules from 2 to 1024 (with a log interval 2) and the policies are random subset of GDPR, HIPAA, FERPA or CCPA following a Gaussian distribution. The experiments are run on a single thread on top of an Ubuntu 16.04 LTS server with 32 AMD Opteron Processors.  The experiments are run for 10 iterations to reach relatively stable results. Performance of \system does not depend on the data, so a real deployment will behave just like the simulation if the policies are similar.

\subsection{Evaluation Results}
\label{sec:eval_results}

In the following, we show and summarize the experiment results. In addition, we discuss and analyze the reasons for our findings.

\begin{wraptable}{r}{0.5\textwidth}
\vspace{-0.2in}
    \centering
    \begin{tabular}{|c|c|c|c|}
    \hline
    Operation & Parsing & Ingestion & Residual Policy \\
    \hline
    Time (ms) & 83 & 9769 & 11\\
    \hline
    \end{tabular}
    \caption{End-to-end evaluation}
    \label{tab:e2e}
    \vspace{-0.2in}
\end{wraptable}

\paragraph{End-to-end evaluation.} We first conduct an end-to-end evaluation to figure out the most time-consuming component in the execution path of \system. We evaluate the time of each component in the execution path in an system with 1024 clients with random subsets of HIPAA.

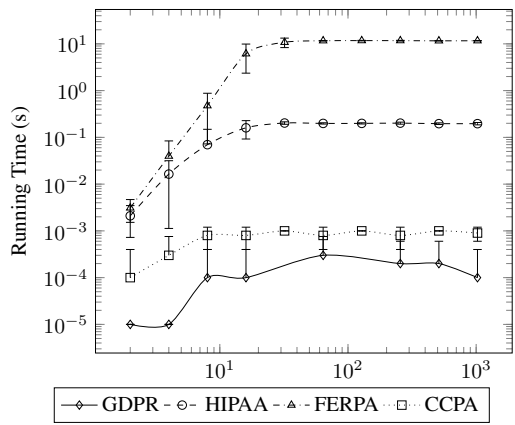
\begin{wrapfigure}{r}{0.5\textwidth}
\begin{tikzpicture}[scale=0.8]
\begin{axis}[legend columns=4,legend style={at={(-0.1,-0.1)},anchor=north west},xmode=log,ymode=log,xlabel=\# Data Capsules,ylabel=Running Time (s)]
\addplot[smooth,mark=diamond,error bars/.cd, y dir=both,y explicit]
table[x=x,y=y,y error=ey]{data/gdpr_lub.dat};
\addlegendentry{GDPR}
\addplot[dashed,smooth,mark=o,mark options={solid},error bars/.cd, y dir=both,y explicit,error bar style={solid}]
table[x=x,y=y,y error=ey]{data/hipaa_lub.dat};
\addlegendentry{HIPAA}
\addplot[dashdotted,smooth,mark=triangle,mark options={solid},error bars/.cd, y dir=both,y explicit,error bar style={solid}]
table[x=x,y=y,y error=ey]{data/ferpa_lub.dat};
\addlegendentry{FERPA}
\addplot[dotted,smooth,mark=square,mark options={solid},error bars/.cd, y dir=both,y explicit,error bar style={solid}]
table[x=x,y=y,y error=ey]{data/ccpa_lub.dat};
\addlegendentry{CCPA}
\end{axis}
\end{tikzpicture}
\caption{Scalability of policy analysis: ingestion operation.}
\label{fig:lub}
\vspace{-0.4in}
\end{wrapfigure}

The results are summarized in \ref{tab:e2e}. The ``Parsing'' column represents the time parsing the analysis program. The ``Ingestion'' column represents the time ingesting the input policies. The ``Residual Policy'' column represents the time computing residual policy given the ingested input policy and the analysis program. Note that the input policy ingestion take up almost all of the running time, which indicates this operation as the bottleneck of the system.

The results demonstrate that the performance overhead of \system is negligible for these programs when the policy guard approach is used, and the bottleneck of \system is the ingestion operation. 

\paragraph{Policy ingestion evaluation}: Next, we perform a targeted microbenchmark to evaluate the scalability of the policy ingestion operation. Figure \ref{fig:lub} contains the results.
As we can observe in the figure, the running time exhibits a polynomial growth (recall that both the x-axis and y-axis are in log scale) at first and then keep stable after the policy number reaches some threshold.
The reason is that the ingestion is implemeted using a least upper bound (LUB) operation, and the LUB operation in \system is composed of two sub-operations: (1) unique with $\mathbb{O}(n\log n)$ complexity, and (2) reduce with $\mathbb{O}(n')$ complexity ($n'$ is the number of policies after unique operation). Because policies are a random subset of some complete policy (GDPR, HIPPA, FERPA and CCPA in this case), if the number of policies are large enough, $n'$ will become a constant. Furthermore the unique operation is $\mathbb{O}(n\log n)$ with a small coefficient so this part is negligible compared to the reduce operation. All these factors result in the trend we observe in Figure \ref{fig:lub}.
This indicates excellent scalability of \system in terms of the number of data capsules.

\section{Related Work}

Recently, there are some research efforts on bootstrapping privacy compliance in big data systems. Technically, the works in this area can be categorized into two directions - (1) summarize the issues in privacy regulations to guide deployment; (2) formalise privacy regulations in a strict programming language flavor; (3) enforce privacy policies in data processing systems. Our work falls into all three categories. In the following, we briefly describe these research works and discuss why these existing approaches do not fully solve the problems in our setting.

\paragraph{Issues in Deploying Privacy Regulations.} 
Gruschka et al.~\cite{gruschka2018privacy} summarize privacy issues in GDPR. Renaud et al.~\cite{renaud2018make} synthesize the GDPR requirements into
a checklist-type format, derive a list of usability design guidelines and providing a usable and GDPR-compliant privacy
policy template for the benefit of policy writers.
Politou et al.~\cite{politou2018forgetting} review all controversies around the new stringent definitions
of consent revocation and the right to be forgotten in GDPR and evaluate existing methods, architectures and
state-of-the-art technologies in terms of fulfilling the technical practicalities for the implementation
and effective integration of the new requirements into current computing infrastructures. Tom et al.~\cite{tom2018conceptual}
present the current state of a model of the GDPR that provides a concise visual overview of the associations between entities defined in the legislation and their constraints.
In this work, our research goal is to summarize and formalize general-purpose privacy principles and design a lightweight paradigm for easy deployment in heterogeneous data processing systems. As a result, these discussions can serve as a good guidance to our work but not actually solve the problem we aim to tackle.
 
\paragraph{Privacy Regulation Formalism.} In \cite{hanson2007data}, Hanson et al. present a data-purpose algebra that can be used to model these kinds of restrictions in various different domains. To formalise purpose restrictions in privacy policies, Tschantz et al.~\cite{tschantz2012formalizing} provide a semantics using a formalism based on planning modeled using a modified version of Markov Decision Processes.
Chowdhury \cite{chowdhury2013privacy} present a policy specification language based on a restricted subset of first order temporal logic (FOTL) which can capture the privacy requirements of HIPAA.
Lam et al.~\cite{lam2012declarative} prove that for any privacy policy that conforms to patterns evident in HIPAA, there exists a finite representative hospital database that illustrates how the law applies in all possible hospitals.
However, because of these works' specific focus on purpose restriction or HIPAA, the above two approaches do not generalize to other regulations like GDPR.
Gerl et al.~\cite{gerl2018lpl} introduce LPL, an extensible Layered Privacy Language that allows to express and enforce these new privacy properties such as personal privacy, user consent, data provenance, and retention management.
Sen et al. introduce \legalease~\cite{sen2014bootstrapping}, a language composed of (alternating) ALLOW and DENY clauses where each clause
relaxes or constricts the enclosing clause. \legalease is compositional and specifies formal semantics in attribute lattices. These characteristics are useful in general-purpose description of privacy regulations and are inherited in \lang. However, compared with \legalease, \lang supports much more expressive attributes to represent abstract domains for static analysis which allows us to encode more complicated privacy regulations. Other work (e.g. Becker et al.~\cite{cassandra}) focuses on the access control issues related to compliance with data privacy regulations, but such approaches do not restrict \emph{how} the data is processed---a key component of recent regulations like GDPR.

\paragraph{Privacy Regulation Compliance Enforcement.}
Going beyond formalism of privacy regulations, recent research also explores techniques to enforce these formalised privacy regulations in real-world data processing systems.
Chowdhury et al.~\cite{chowdhury2014temporal} propose to use temporal model-checking for run-time monitoring of privacy policies. While Chowdhury demonstrates the effectiveness of this approach in online monitoring of privacy policies, it does not provide the capability of static analysis to decide if a analytic program satisfies a privacy policy and can only report privacy violation after it happens.
Sen et al.~\cite{sen2014bootstrapping} introduce {\sc Grok}, a data inventory for Map-Reduce-like big data systems. Although working perfectly in Map-Reduce-like systems, {\sc Grok} lacks adaptability to non-Map-Reduce-like data processing systems.


\vspace{-0.15in}
\section{Conclusion \& Future Work}
\vspace{-0.1in}

In this paper, we have proposed the data capsule paradigm, a new paradigm for collecting, managing, and processing sensitive personal data. The data capsule paradigm has the potential to break down data silos and make data more useful, while at the same time reducing the prevalence of data privacy violations and making compliance with privacy regulations easier for organizations. We implemented \system, a reference platform for the new paradigm.


We are currently in the preliminary stages of a collaborative case study to apply the data capsule paradigm to enforce HIPAA in a medical study of menstrual data collected via mobile app. The goal of this study~\cite{Symul385054} and similar work~\cite{liu2018predicting,10.1093/emph/eoy018} is to demonstrate the use of mobile apps to assess menstrual health and fertility. Data capsules will allow study participants to submit their sensitive data in the context of a policy which protects its use. As part of this effort, we are in the process of encoding the requirements of HIPAA using \lang and applying \system to the analysis programs written by the study's designers.

\section*{Acknowledgments}

We thank the anonymous reviewers for their helpful comments. 
This work was supported by DARPA \& SPAWAR under contract N66001-15-C-4066. The U.S. Government is authorized to reproduce and distribute reprints for Governmental purposes not withstanding any copyright notation thereon. The views, opinions, and/or findings expressed are those of the author(s) and should not be interpreted as representing the official views or policies of the Department of Defense or the U.S. Government.

\bibliographystyle{IEEEtran}
\bibliography{ref}

\begin{thebibliography}{10}
\providecommand{\url}[1]{#1}
\csname url@samestyle\endcsname
\providecommand{\newblock}{\relax}
\providecommand{\bibinfo}[2]{#2}
\providecommand{\BIBentrySTDinterwordspacing}{\spaceskip=0pt\relax}
\providecommand{\BIBentryALTinterwordstretchfactor}{4}
\providecommand{\BIBentryALTinterwordspacing}{\spaceskip=\fontdimen2\font plus
\BIBentryALTinterwordstretchfactor\fontdimen3\font minus
  \fontdimen4\font\relax}
\providecommand{\BIBforeignlanguage}[2]{{%
\expandafter\ifx\csname l@#1\endcsname\relax
\typeout{** WARNING: IEEEtran.bst: No hyphenation pattern has been}%
\typeout{** loaded for the language `#1'. Using the pattern for}%
\typeout{** the default language instead.}%
\else
\language=\csname l@#1\endcsname
\fi
#2}}
\providecommand{\BIBdecl}{\relax}
\BIBdecl

\bibitem{databreach}
``The 18 biggest data breaches of the 21st century,''
  \url{https://www.csoonline.com/article/2130877/the-biggest-data-breaches-of-the-21st-century.html},
  2019, online; accessed 23 May 2019.

\bibitem{solove2017risk}
D.~J. Solove and D.~K. Citron, ``Risk and anxiety: A theory of data-breach
  harms,'' \emph{Tex. L. Rev.}, vol.~96, p. 737, 2017.

\bibitem{insiderattack}
``Insider threat 2018 report,''
  \url{https://www.ca.com/content/dam/ca/us/files/ebook/insider-threat-report.pdf},
  2019, online; accessed 23 May 2019.

\bibitem{murdock1979use}
L.~E. Murdock, ``The use and abuse of computerized information: Striking a
  balance between personal privacy interests and organizational information
  needs,'' \emph{Alb. L. Rev.}, vol.~44, p. 589, 1979.

\bibitem{GDPR}
``The eu general data protection regulation (gdpr),''
  \url{https://eugdpr.org/}, 2019, online; accessed 16 April 2019.

\bibitem{CCPA}
``California consumer privacy act (ccpa),'' \url{https://www.caprivacy.org/},
  2019, online; accessed 16 April 2019.

\bibitem{FERPA}
``The family educational rights and privacy act of 1974 (ferpa),''
  \url{https://www.colorado.edu/registrar/students/records/ferpa}, 2019,
  online; accessed 16 April 2019.

\bibitem{HIPAA}
``Health insurance portability and accountability act (hipaa),''
  \url{https://searchhealthit.techtarget.com/definition/HIPAA}, 2019, online;
  accessed 16 April 2019.

\bibitem{forever}
``Google keeps your data forever - unlocking the future transparency of your
  past,''
  \url{https://www.siliconvalleywatcher.com/google-keeps-your-data-forever---unlocking-the-future-transparency-of-your-past/},
  2019, online; accessed 30 May 2019.

\bibitem{maniatis2011you}
P.~Maniatis, D.~Akhawe, K.~R. Fall, E.~Shi, and D.~Song, ``Do you know where
  your data are? secure data capsules for deployable data protection.'' in
  \emph{HotOS}, vol.~7, 2011, pp. 193--205.

\bibitem{ETL}
``Extract, transform, load,''
  \url{https://en.wikipedia.org/wiki/Extract,_transform,_load}, 2019, online;
  accessed 30 May 2019.

\bibitem{sql}
E.~F. Codd, ``A relational model of data for large shared data banks,''
  \emph{Communications of the ACM}, vol.~13, no.~6, pp. 377--387, 1970.

\bibitem{mongodb}
K.~Chodorow, \emph{MongoDB: the definitive guide: powerful and scalable data
  storage}.\hskip 1em plus 0.5em minus 0.4em\relax " O'Reilly Media, Inc.",
  2013.

\bibitem{hdfs}
K.~Shvachko, H.~Kuang, S.~Radia, R.~Chansler \emph{et~al.}, ``The hadoop
  distributed file system.'' in \emph{MSST}, vol.~10, 2010, pp. 1--10.

\bibitem{cassandra}
A.~Lakshman and P.~Malik, ``Cassandra: a decentralized structured storage
  system,'' \emph{ACM SIGOPS Operating Systems Review}, vol.~44, no.~2, pp.
  35--40, 2010.

\bibitem{mapreduce}
J.~Dean and S.~Ghemawat, ``Mapreduce: simplified data processing on large
  clusters,'' \emph{Communications of the ACM}, vol.~51, no.~1, pp. 107--113,
  2008.

\bibitem{hadoop}
K.~Shvachko, H.~Kuang, S.~Radia, R.~Chansler \emph{et~al.}, ``The hadoop
  distributed file system.'' in \emph{MSST}, vol.~10, 2010, pp. 1--10.

\bibitem{spark}
M.~Zaharia, M.~Chowdhury, M.~J. Franklin, S.~Shenker, and I.~Stoica, ``Spark:
  Cluster computing with working sets.'' \emph{HotCloud}, vol.~10, no. 10-10,
  p.~95, 2010.

\bibitem{sen2014bootstrapping}
S.~Sen, S.~Guha, A.~Datta, S.~K. Rajamani, J.~Tsai, and J.~M. Wing,
  ``Bootstrapping privacy compliance in big data systems,'' in \emph{2014 IEEE
  Symposium on Security and Privacy}.\hskip 1em plus 0.5em minus 0.4em\relax
  IEEE, 2014, pp. 327--342.

\bibitem{conceptlattice}
``Formal concept analysis,''
  \url{https://en.wikipedia.org/wiki/Formal_concept_analysis}, 2019, online;
  accessed 30 May 2019.

\bibitem{nielson2015principles}
F.~Nielson, H.~R. Nielson, and C.~Hankin, \emph{Principles of program
  analysis}.\hskip 1em plus 0.5em minus 0.4em\relax Springer, 2015.

\bibitem{gruschka2018privacy}
N.~Gruschka, V.~Mavroeidis, K.~Vishi, and M.~Jensen, ``Privacy issues and data
  protection in big data: A case study analysis under gdpr,'' in \emph{2018
  IEEE International Conference on Big Data (Big Data)}.\hskip 1em plus 0.5em
  minus 0.4em\relax IEEE, 2018, pp. 5027--5033.

\bibitem{renaud2018make}
K.~Renaud and L.~A. Shepherd, ``How to make privacy policies both
  gdpr-compliant and usable,'' in \emph{2018 International Conference On Cyber
  Situational Awareness, Data Analytics And Assessment (Cyber SA)}.\hskip 1em
  plus 0.5em minus 0.4em\relax IEEE, 2018, pp. 1--8.

\bibitem{politou2018forgetting}
E.~Politou, E.~Alepis, and C.~Patsakis, ``Forgetting personal data and revoking
  consent under the gdpr: Challenges and proposed solutions,'' \emph{Journal of
  Cybersecurity}, vol.~4, no.~1, p. tyy001, 2018.

\bibitem{tom2018conceptual}
J.~Tom, E.~Sing, and R.~Matulevi{\v{c}}ius, ``Conceptual representation of the
  gdpr: Model and application directions,'' in \emph{International Conference
  on Business Informatics Research}.\hskip 1em plus 0.5em minus 0.4em\relax
  Springer, 2018, pp. 18--28.

\bibitem{hanson2007data}
C.~Hanson, T.~Berners-Lee, L.~Kagal, G.~J. Sussman, and D.~Weitzner,
  ``Data-purpose algebra: Modeling data usage policies,'' in \emph{Eighth IEEE
  International Workshop on Policies for Distributed Systems and Networks
  (POLICY'07)}.\hskip 1em plus 0.5em minus 0.4em\relax IEEE, 2007, pp.
  173--177.

\bibitem{tschantz2012formalizing}
M.~C. Tschantz, A.~Datta, and J.~M. Wing, ``Formalizing and enforcing purpose
  restrictions in privacy policies,'' in \emph{2012 IEEE Symposium on Security
  and Privacy}.\hskip 1em plus 0.5em minus 0.4em\relax IEEE, 2012, pp.
  176--190.

\bibitem{chowdhury2013privacy}
O.~Chowdhury, A.~Gampe, J.~Niu, J.~von Ronne, J.~Bennatt, A.~Datta, L.~Jia, and
  W.~H. Winsborough, ``Privacy promises that can be kept: a policy analysis
  method with application to the hipaa privacy rule,'' in \emph{Proceedings of
  the 18th ACM symposium on Access control models and technologies}.\hskip 1em
  plus 0.5em minus 0.4em\relax ACM, 2013, pp. 3--14.

\bibitem{lam2012declarative}
P.~E. Lam, J.~C. Mitchell, A.~Scedrov, S.~Sundaram, and F.~Wang, ``Declarative
  privacy policy: finite models and attribute-based encryption,'' in
  \emph{Proceedings of the 2nd ACM SIGHIT International Health Informatics
  Symposium}.\hskip 1em plus 0.5em minus 0.4em\relax ACM, 2012, pp. 323--332.

\bibitem{gerl2018lpl}
A.~Gerl, N.~Bennani, H.~Kosch, and L.~Brunie, ``Lpl, towards a gdpr-compliant
  privacy language: Formal definition and usage,'' in \emph{Transactions on
  Large-Scale Data-and Knowledge-Centered Systems XXXVII}.\hskip 1em plus 0.5em
  minus 0.4em\relax Springer, 2018, pp. 41--80.

\bibitem{chowdhury2014temporal}
O.~Chowdhury, L.~Jia, D.~Garg, and A.~Datta, ``Temporal mode-checking for
  runtime monitoring of privacy policies,'' in \emph{International Conference
  on Computer Aided Verification}.\hskip 1em plus 0.5em minus 0.4em\relax
  Springer, 2014, pp. 131--149.

\bibitem{Symul385054}
\BIBentryALTinterwordspacing
L.~Symul, K.~Wac, P.~Hillard, and M.~Salathe, ``Assessment of menstrual health
  status and evolution through mobile apps for fertility awareness,''
  \emph{bioRxiv}, 2019. [Online]. Available:
  \url{https://www.biorxiv.org/content/early/2019/01/28/385054}
\BIBentrySTDinterwordspacing

\bibitem{liu2018predicting}
B.~Liu, S.~Shi, Y.~Wu, D.~Thomas, L.~Symul, E.~Pierson, and J.~Leskovec,
  ``Predicting pregnancy using large-scale data from a women's health tracking
  mobile application,'' \emph{arXiv preprint arXiv:1812.02222}, 2018.

\bibitem{10.1093/emph/eoy018}
\BIBentryALTinterwordspacing
A.~Alvergne, M.~Vlajic~Wheeler, and V.~Högqvist~Tabor, ``{Do sexually
  transmitted infections exacerbate negative premenstrual symptoms? Insights
  from digital health},'' \emph{Evolution, Medicine, and Public Health}, vol.
  2018, no.~1, pp. 138--150, 07 2018. [Online]. Available:
  \url{https://doi.org/10.1093/emph/eoy018}
\BIBentrySTDinterwordspacing

\end{thebibliography}

\end{document}